# Improved Service Delivery and Cost Effective Framework for e-Governance in India

Puneet Kumar
Assistant Professor
MITS University
Lakshmangarh, Sikar,
Rajasthan
India

Dharminder Kumar
Professor & Chairman
Guru Jambheshwar University
of Science & Technology
Hissar, Haryana
India

Narendra Kumar
Assistant Professor
MITS University
Lakshmangarh, Sikar
Rajasthan
India

## ABSTRACT

In current era, the involvement of technologies like virtualization, consolidation and cloud computing, and adoption of free and open source software in designing and deploying e-governance that can reduce the total cost associated with and hence the financial burden abide by the state and central governments. The success of any e-governance project depends upon its utilization by the intended group and so there accessibility needs to be enhanced drastically by reengineered framework. Here, we design an Improved Service Delivery and Cost Effective Framework for e-Governance that will be useful for success of e-governance projects and the delivery mechanism in India by using free and open access software for development and deployment of e-governance applications, virtualization and consolidation techniques for management of e-services and cloud computing for enhancing the accessibility of services.

## 1. INTRODUCTION

Targeting the Indian population over 1.2 billion for delivering electronic services (e-services) is mammoth task and requires adequate implementation policies [1]. Since, Information and Communication Technology (ICT) is playing a significant role in delivery of e-services up to grass root level therefore Information Technology (IT) resources needs to be managed prudently in terms of incurred cost and service delivery. The 'true' value of electronic connectivity is a function of its relevance and user-friendliness:

**Value of electronic connectivity=Relevance x Usability**

Relevance is the balancing factor between technology available and the cost economics [2]. Government of India is following two models for imparting e-governance services viz NIC (National Informatics Centre) model and PPP (Public Private Partnership) model [3]. For availing e-services, users have to pay the requisite amount to either government of private partners. According to World Bank, 81.1% (992 million) of the Indian population is living on less than $2.5 (INR 130 approx.) per day [4].In such type of circumstances where marginalized group is dominating, the e-governance implementation mechanism requires reengineering in such a way that the marginalized groups can also be benefitted to great extent. In this paper we are proposing an abstract framework and some methodologies which can help in

reducing IT infrastructure costs and enhancement in the delivery of e-services.

## 2. REVIEW OF LITERATURE

In India 35% of the e-governance projects are total failures because of the reason that initiatives are not implemented or abandoned immediately. The 50% of the projects are partial failures because they are not able to achieve the supposed objectives and only 15% of the e-governance projects attain success in all the respects. The main reasons behind the failure of projects are inadequate project definition, deficiency of change control system, poor project estimations, lack of skills and ineffective resource management [5][6]. There are various factors which are responsible for the failure of e-governance projects in India. Improper planning, incorrect cost estimations, vendor lock in and hustle in implementation of any e-governance projects are the major factors which lead to the failure [7]. The Government Data Centre (GDC) is the most important component of e-governance infrastructure. The challenges occurred in the designing of a GDC are higher capital cost for infrastructure, higher operating costs in managing and administering various types of servers, storage utilization and recurring deployment of multiple applications on distinct platforms [8]. Since the developing countries are facing financial crises of funds for performing heavy expenditure on latest technology, therefore, Beowulf cluster comprises of obsolete machines can be used in low budget regions. It can be used as method to bridge the digital divide and ensure the accessibility of computers to disadvantaged group. Beowulf cluster method can be used to enhance the accessibility of e-services but it is imperfect for server administration and hosting [9].e-Governance is facing a problem of unproductive investments and it needs to be monitored [10]. To achieve the objectives of National e-Governance Plan (NeGP), the service delivery mechanism needs to be improved and grid technology can play a vital role in this regard [11]. Virtualization and cloud computing is a cost effective solution for delivering e-governance services to the citizens. Nevertheless there are some security issues but despite of it is capable to ensure faster and cheaper delivery of services through service oriented architecture [12].

## 3. GOVERNMENT DATA CENTERS (GDCs): THE CURRENT SCENARIO

The GDC is the main important component of information and communication technology (ICT) infrastructure for supporting e-governance initiatives. The key role of GDC is to amalgamate services and infrastructure to disseminate various

Government to Government (G2G), Government to Business (G2B), Government to Citizen (G2C) and Government to Employee (G2E) services. A GDC behaves as a central repository of the state for secure data storage, online delivery of e-services, management and maintenance of citizen centered portals, disaster recovery mechanism, remote management and service integration.

Currently, GDCs encompasses individual servers for separate applications where each server requires its own power along with air conditioning and real estate rack space. Since GDCs have an inclination towards physical growth beyond the expectations therefore it leads towards installation of additional servers, storage and physical space [13]. The major challenges faced by GDCs are mentioned us under:

3.1 Higher capital costs for infrastructure which includes ICT infrastructure and real estate infrastructure

3.2 Higher operational costs for managing and administering various types of servers and operating environments

3.3 Elevated power and cooling requirements

3.4 Optimal storage utilization

3.5 Deployment and re-deployment of multiple applications on different platforms

## 4. FRAMEWORK

The proposed framework involves inculcation of virtualization and consolidation techniques, use of Free and Open Source Software (FOSS) and utility computing together for enhancing service delivery and reducing the expenditure incurred in facilitating e-services.

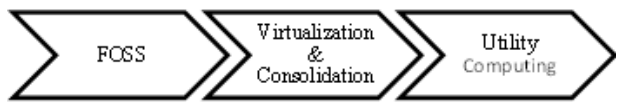

Fig.1: Steps towards cost effective framework

## 4.1 Free and Open Source Software (FOSS)

A FOSS is abundantly licensed to the user and it also grants the rights associated with the software viz right to use, copy, study, change, and improve its design through the availability of its source code [14]. The use of FOSS reduces the total cost of ownership and vendor locking. The open source software needs to be standardized because of the involvement of national security. After standardization the open source software should be used nationwide for the development and deployment of e-governance applications [15]. Without adoption of FOSS, the government is bound to obtain various types of software licenses as well as licenses for terminals from private vendors and India is spending an enormous amount for the same every year. The IT infrastructure cost is one of the major hurdles in successful implementation of e-governance in developing nations [16]. Therefore the use of FOSS must be encouraged and it must be standardized in such a way that it must have scalability and interoperability so that different applications can communicate with each other for reducing the redundancy of data.

## 4.2 Virtualization & Consolidation

Virtualization is a technique which provides a layer of abstraction between computing, storage and networking infrastructure and applications running on it whereas consolidation consolidates variety of servers together [17]. Consolidation reduces the data centers complexities, maintenance cost and energyconsumptions. It enables efficient utilization of hardware resources in order to reduce total number of servers and server locations [18]. To deal with the aforesaid challenges in implementation of a GDC, virtualization can play a pivotal role at software as well as hardware level. There are various vendors like SUN, IBM, HP etc. having variety of virtualization techniques. Virtualization techniques can be classified in to various types of categories mentioned as under:

### 4.2.1 Server virtualization

It deals with virtualizing servers at hardware level to enhance flexibility and availability.

### 4.2.2 Operating system virtualization

It enables the users to run different kinds of operating systems and applications on the same set of hardware in isolation with each other.

### 4.2.3 Storage virtualization

It consolidates heterogeneous storages together and represents as a common storage pool of resources. The virtualization software ensures the availability of relevant data to the users without revealing its location.

### 4.2.4 Network virtualization

The network switches can be virtualized as Virtual Local Area Networks (VLANs) in secure isolation with each other.

### 4.2.5 Client virtualization

It enables the user to access any application from any mode of communication without making any modification in the application.In nutshell virtualization techniques are available to manage and provide a variety of servers and operating environments.Consolidation is the technique which helps in making efficient use of resources by increasing system utilization, lowering Total Cost of Ownership (TCO), improving profitability and enhancement in the ability to respond to citizens. It provides the facility to run multiple applications, like web server and database server on multiple virtual partitions while sharing the same physical hardware. It leads towards reduction in TCO, easy management and savings in power and space requirements [19].

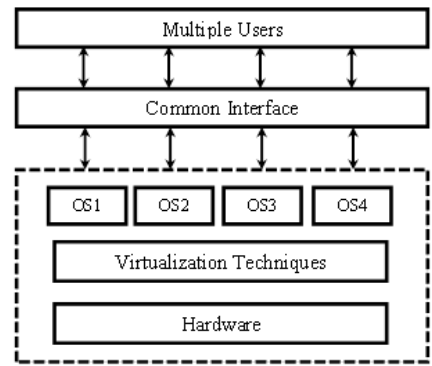

Fig. 2: Virtualization and Consolidation (Bhattacharya, 2012)

## 4.3 Utility Computing

It is a way by which an IT service provider provides IT services in infrastructure, applications and business areas on need basis and charges the services on specific usages basis. It is also referred as cloud computing. The cloud computing can reduce the infrastructure cost to some extent because it imparts various types of services like Infrastructure as a Service (IaaS), Platform as aService (PaaS) and Software as a Service (SaaS) to enhance flexibilityin network operations and data operations among network partners through service oriented architecture [20]. In India 72% of the populace is having cell phones [21], therefore to enhance the accessibility of e-services especially in remote areas cloud computing can be used as a pivot. It enables the recipients to access the services through cellular phones also. The government should design a G-Cloud (Government Cloud) for offering all kinds of e-services. It will overcome various challenges pertaining to accessibility of e-services in remote areas like connectivity, power and cooling infrastructure and IT infrastructure.

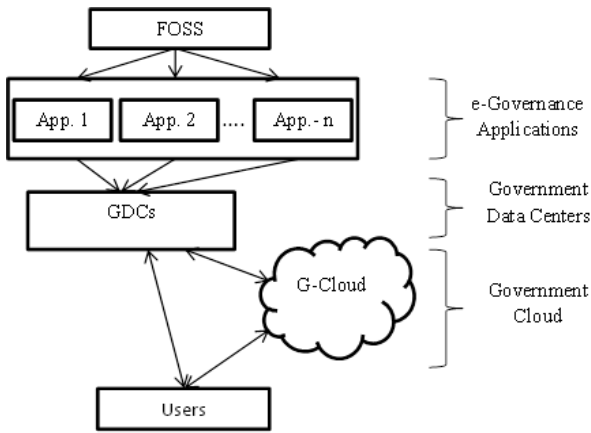

Fig.3: Framework for cost effective e-governance

## 5. FEATURES

The proposed framework can provide a cost effective solution for e-governance by using Free and Open Access Software for development and deployment of e-governance applications, virtualization and consolidation techniques for management of e-services and cloud computing for enhancing the accessibility of services among citizens including rural masses. The major features of the proposed framework are depicted are under:

- It will help in reducing the cost of hardware incurred in facilitating e-governance services to citizens.
- It helps in reducing budding real estate as well as cooling requirements in establishment as well as maintenance of GDCs.
- It also reduces the total cost of ownership on e-governance projects by reducing the costs incurred on acquiring various types of hardware and software licenses from private vendors.
- It also gives solution to enhance the availability and accessibility of e-services in disadvantaged areas of country.
- It facilitates scalability of e-governance projects as the data volume in e-governance projects increases enormously.

- It promotes interoperability in the e-governance applications for reducing data redundancy and hence increases consistency of data.

## 6. CASE STUDIES

- The Japan government has devised a G-Cloud named as "Kasumigaseki Cloud" to host all Japanese government's computing. It allows greater information and resources sharing by utilizing virtualization and consolidation techniques in government's IT resources. It not only cost effective but environment friendly also. In Thailand, Government Information Technology Service (GITS) has established a cloud for serving all Thai government agencies. The Vietnam government is also working with IBM to formulate a cloud to serve public and private sectors. China has established the Yellow river Delta Cloud Computing Center not only to improve only e-governance service delivery but also for the economic development of the country [22].
- In Sweden a company Hi3G Access AB were having challenges like reducing infrastructure costs, improving management of data centre, improving scalability and performance of the application. To conquer these challenges the company has adopted virtualized environment and as a result the total number of servers were reduced from 400 to 75 and floor space cut by 75%. The speed of the application was also enhanced drastically; indeed there was a great reduction in IT infrastructure as well as real estate requirements [23]. iTricity company in Netherlands has reduced their asset utilization by 25% to 30% and operational costs by 30 % to 35 % in last three years by adopting virtualization [24] whereas EDIF of Italy had reduced their operational costs by 75% [25]. Le Bourget Communauté of France has saved 20% of the total cost along with continuous savings in power consumption [26] and Denver Health Medical Center of Denver has also reduced their operational costs along with a significant improvement in service delivery by espousing virtualization [27].
- In Middle East and African (MEA) countries the growth of x86 servers is declining by 5.9% year by year with a significant growth in revenue by 3.0% whereas the shipment of blade servers has increased up to 10.9% in the MEA region. It shows the deployment of virtualization in business enterprises [28].

## 7. CONCLUSION

The involvement of technologies like virtualization, consolidation and cloud computing and adoption of free and open source software in designing and deploying e-governance will lead towards reduction in total cost associated with both hardware as well as software. Therefore it reduces the financial burden abide by the state and central governments. For ensuring the successfulness of e-governance projects the delivery mechanism needs to be reengineered. The success of any e-governance project depends upon its utilization by the intended group and hence there accessibility needs to be enhanced drastically. This paper suggests a deigned framework for improved and cost effective e-governance schemes in India.